# A VM-HDL Co-Simulation Framework for Systems with PCIe-Connected FPGAs

Shenghsun Cho*, Mrunal Patel*, Basavaraj Kaladagi*,
Han Chen†, Tapti Palit*, Michael Ferdman*, Peter Milder†
*Department of Computer Science
†Department of Electrical and Computer Engineering
Stony Brook University
Stony Brook, NY 11794

*Abstract*—PCIe-connected FPGAs are gaining popularity as an accelerator technology in data centers. However, it is challenging to jointly develop and debug host software and FPGA hardware. Changes to the hardware design require a time-consuming FPGA synthesis process, and modification to the software, especially the operating system and device drivers, can frequently cause the system to hang, without providing enough information for debugging. The combination of these problems results in long debug iterations and a slow development process. To overcome these problems, we designed a VM-HDL co-simulation framework, which is capable of running the same software, operating system, and hardware designs as the target physical system, while providing full visibility and significantly shorter debug iterations.

## I. INTRODUCTION

FPGAs are gaining popularity as an accelerator technology to offload complex computation and data flows. The combination of programmability, high degree of parallelism, and low power consumption make FPGAs suitable for environments with rapidly changing workloads and strict power consumption limits, such as data centers. To put FPGAs into existing systems, PCIe has become the most common connection choice, due to its wide availability in server systems. Today, the majority of FPGAs in data centers are communicating with the host system through PCIe [1], [2].

Unfortunately, developing applications for FPGAs requires the time-consuming FPGA compilation processes, including synthesis, place, and route. Moreover, it is challenging to develop and debug the host software and the FPGA hardware designs at the same time. The hardware designs running on the FPGAs provide little to no visibility, and even small changes to the hardware may need hours to go through the FPGA compilation process. The development process becomes even more difficult when taking operating systems into account. Changes to the operating system kernel, the loadable kernel modules, and the application software and hardware can frequently hang the system without providing enough information for debug, forcing a tedious reboot process. The combination of these problems results in long debug iterations and a slow development process.

The traditional way to test and debug hardware designs without running them on real FPGAs is by writing testbenches for simulation. The main drawback of this approach is that the hardware cannot be tested together with the software and operating system. While some vendors provide hardware-software co-simulation environments, they still lack the ability to cover the development of operating system and device driver code. There are also frameworks that connect an instruction-set simulator to an HDL simulator to perform full-system simulation. However, these frameworks target system-on-chips—typically, ASICs with ARM cores—and thus cannot simulate the PCIe-connected FPGAs used in data center servers.

We observe that, although there is no readily available environment for full-system simulation of servers with PCIe-connected FPGAs, we can extend existing tools to build a co-simulation framework for debugging such systems. Virtual machines (VMs) are widely used to run services in data centers. The capability of emulating a full system, including CPUs, disks, memory, and peripherals such as PCIe devices, makes VMs a natural fit to emulate the server system in a development environment. On the other hand, FPGA vendors are providing sophisticated software for developers to generate FPGA platforms with PCIe interfaces for real hardware and for HDL simulators with testbenches. The key missing component is a link between a VM's virtual PCIe device and the PCIe block in an FPGA HDL simulation platform.

In this work, we developed a co-simulation framework using communication channels between a VM and an HDL simulator. On the VM side, we created a PCIe FPGA pseudo device to represent the FPGA board. The operating system and software running inside the VM see the same PCIe device as if they were running in a real system with an FPGA board plugged in. On the HDL side, we developed a PCIe simulation bridge to talk to the VM. The PCIe simulation bridge is pin-compatible with the PCIe block used in the physical FPGA hardware. The rest of the FPGA platform sees the same interface toward PCIe and requires no modification. To the FPGA development tools, the PCIe simulation bridge appears as a regular hardware block and has no impact on the simulation flow. Notably, we linked the VM's PCIe FPGA pseudo device and the PCIe simulation bridge together using a high-level queue library that provides reliable message passing, which has an additional useful benefit: either side of the simulation can be independently restarted without affecting the other side.

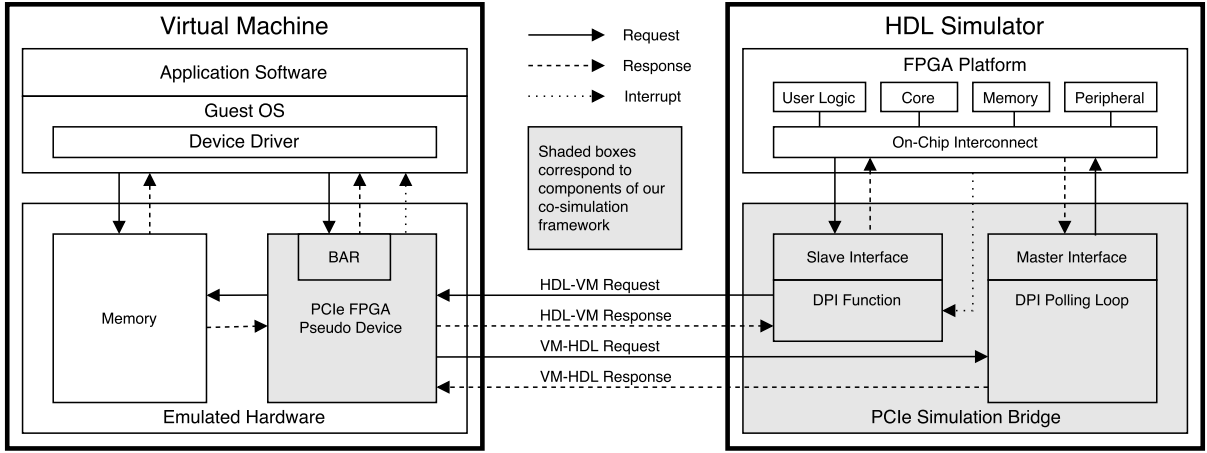

Fig. 1. The VM-HDL co-simulation framework

To demonstrate and evaluate our VM-HDL co-simulation framework, we built a sorting offload design. Our experiments with this design indicate that the co-simulation framework significantly reduces the debug iteration time. Moreover, our framework provides invaluable visibility into both the hardware design and the operating system, making it easier and faster to identify problems while developing and debugging.

## II. THE VM-HDL CO-SIMULATION FRAMEWORK

The key component in our co-simulation framework is the link between the virtual machine monitor (VMM) and the HDL simulator. The most important requirement for this link is to expose exactly the same interface and functionality as its counterpart in the real system, to allow a smooth transition between the development and debugging of the system in the co-simulation framework and its deployment on real hardware. All parts of the system, including the FPGA platform, the operating system driver, and the software, must be able to run in simulation and on the target hardware without any modifications. Another requirement is that the cut-off point of the link and the rest of the system should be generic and well-defined, to reduce the effort of developing and using the link. With these considerations in mind, we developed a VM-HDL link with three parts: a PCIe FPGA pseudo device, a PCIe simulation bridge, and message passing channels between them. The architecture of the co-simulation framework is shown in Figure 1.

We created a PCIe FPGA pseudo device in the VMM to represent the PCIe FPGA board. The structure of the VMM's emulated PCIe devices is generic and well defined, enabling us to create the PCIe FPGA pseudo device by modifying an existing device and customizing it with the target FPGA board's PCIe characteristics, such as the number and size of the Base Address Register (BAR) regions and Message Signaled Interrupt (MSI) capabilities. The benefit of using a PCIe pseudo device is that it interacts with the guest operating system through Memory-Mapped Input-Output (MMIO), thus allowing us to avoid the low-level PCIe protocol details. MMIO read and write requests to the BAR regions are handled using callback functions and translated into messages that are sent to the HDL simulator. The PCIe FPGA pseudo device also configures the VMM to listen to memory accesses and interrupts from the HDL side by registering the file descriptors of the communication channels with the VMM's main loop, enabling the VMM subsystem to respond to the memory read, write, and interrupt requests from the HDL simulator.

On the HDL side, we developed a PCIe simulation bridge to replace the hardware PCIe bridge in the FPGA platform. To avoid implementing the low-level PCIe protocol, we rely on an industry-standard on-chip bus protocol, the Advanced eXtensible Interface (AXI). AXI serves as the bridge's interface to the rest of the FPGA platform. A slave interface monitors the AXI bus signals for memory access requests to the simulation bridge, which triggers the corresponding functions, implemented using SystemVerilog DPI, to send these requests to the VMM. The simulation bridge also listens to requests and reads responses from the VMM, calling the corresponding HDL tasks to either send MMIO read and write requests to the FPGA platform through the AXI master interface, or to send back read responses to the FPGA platform through the AXI slave interface. An interrupt pin on the simulation bridge's interface allows the FPGA platform to also send requests that generate MSI interrupts in the VM.

The VMM's PCIe FPGA pseudo device and the PCIe simulation bridge communicate through two pairs of unidirectional channels, one for HDL to VMM accesses and the other for VMM to HDL accesses. We use a high-level queue library that provides reliable message passing. In each of the channel pairs, one channel is used to send requests and the other is used to receive responses. Using multiple unidirectional channels provides the necessary independence between the VM and the HDL simulator to allow rebooting/restarting either side without affecting the other. The channels carry messages that contain the request and response information such as address, length, and data. The structure of the messages can be easily extended to carry additional customized information.

TABLE I
SETUP FOR CO-SIMULATION AND PHYSICAL SYSTEM

| | |
|---|---|
| Target FPGA Board | NetFPGA SUME (xc7vx690tffg1761-3) |
| Co-Sim Host Hardware | Xeon E5-2620 v4 with 64GB DDR4 |
| Co-Sim Host OS | Ubuntu 14.04 with Kernel 3.13.0 and KVM |
| VMM | QEMU 2.7.50 |
| FPGA Tool | Xilinx Vivado 2016.2 |
| HDL Simulator | Synopsys VCS J-2014.12-SP3-8 |
| Message Passing Library | ZeroMQ 4.2.1 |
| FPGA Compilation Host | Xeon E5-2620 v3 with 64GB DDR4 |
| Physical System Hardware | Xeon E5-2620 v3 with 64GB DDR4 |
| Physical System OS | Ubuntu 14.04 with Kernel 3.16.7 |

TABLE II
RUN TIME COMPARISON

| | Physical System (sec.) | Co-Simulation (sec.) |
|---|---|---|
| Compilation | - | 167 |
| Synthesis | 1617 | - |
| Place and Route | 2672 | - |
| Reboot | 120 | - |
| Execution | 0.000032 | 6.02 |
| Total | ≈4409 | ≈173 |

By using the co-simulation framework, the guest operating system inside the VM interacts with the HDL simulator as it would with a physical PCIe FPGA, and the FPGA platform interacts with the VM as it would with a physical host system. The software, operating system, device driver, and FPGA platform remain unmodified between the co-simulation and real hardware environments, eliminating the porting effort from one to the other. In terms of debugging, in addition to debugging software by running GDB in the guest OS, our co-simulation framework allows developers to connect GDB to the VMM's debugging interface to debug the operating system and device driver code, enabling advanced functionality such as single-stepping kernel instructions, including inside interrupt handlers, and monitoring or even modifying register and memory contents. On the FPGA side, unlike in a logic analyzer-like environment that limits the number of probed signals and requires re-synthesis to insert additional probes, developers can record signals of the entire FPGA platform during the entire simulation and even force signal values, providing full visibility for debugging and analysis.

## III. IMPLEMENTATION AND EVALUATION

To demonstrate our VM-HDL co-simulation framework, we developed an FPGA-based sorting offload platform. The sorting unit we used in the platform is automatically generated by the Spiral Sorting Network IP Generator [3]. The sorting unit takes a stream of input data and produces the output result stream after a fixed number of cycles. The sorting unit is fully pipelined and able to consume back-to-back input streams.

We build the FPGA platform using Xilinx Vivado 2016.2, targeting the NetFPGA SUME PCIe board. The sorting unit in our accelerator uses 128-bit wide stream interfaces for input and output and can sort 1024 32-bit signed integers in 1256 cycles. A Xilinx DMA is used to fetch input data from the host memory through PCIe, stream data through the sorting unit, and write the results back to the host memory. For co-simulation, we replace the Xilinx PCIe-AXI bridge with our PCIe simulation bridge. We use QEMU with KVM support as the VMM and use ZeroMQ [4] as the queue library to link the VM and the HDL simulator together. While performing co-simulation, we record waveforms in the FSDB format for the entire FPGA platform. For comparison, we run the design on the physical FPGA platform, using the Xilinx PCIe-AXI bridge. The FPGA LUT utilization after place and route is 11%, and the BRAM utilization is 19%.

The setup of the hardware and software for co-simulation and physical system are shown in Table I.

## IV. RESULTS

### A. Debug Iteration Time

The primary advantage of our co-simulation framework is the improvement in the debug iteration time, the time needed to make a change to the software or hardware description and observe its results. For physical systems, this is particularly noticeable when the system "hangs" due to a bug and requires a tedious reboot or when changes to the FPGA platform hardware require the time-consuming FPGA compilation process. Even worse, the physical system does not provide sufficient visibility into the hardware or software when an error occurs, requiring developers to go through many debug iterations to find and fix each bug. To contrast the two debug approaches, Table II shows the breakdown of the debug iteration time on a physical system and on our co-simulation framework. Compared to a physical system, which requires the FPGA compilation process, the co-simulation framework is 25x faster in our test case. Changes in the FPGA platform can run in the HDL simulator with full visibility in just a few minutes.

### B. Application Execution Time

Although the debug iteration time is drastically reduced, the simulation platform runs slower than the physical system, as shown in Table II. This is expected because the co-simulation framework performs cycle-accurate HDL simulation and, on every cycle, the simulator performs additional work to poll the communication channels to see if there is any request from the VM. The results suggest that developers should avoid long test cases while debugging using the co-simulation platform, which is true for any HDL simulation environment. However, with relatively small test cases, the execution time is acceptable for interactive debugging and iteration, and the time savings from shorter debug iterations and greater visibility into the design easily mitigate the longer application execution times.

TABLE III
COMPARISON BETWEEN ACTUAL TIME AND SIMULATED TIME

|  | Actual Time ($\mu$s) | Simulated Time ($\mu$s) |
| --- | --- | --- |
| Host to Device Read RTT | 0.85 | 72,400 |
| Application Execution Time | 32 | 6,023,300 |

*C. Simulated Time*

Our co-simulation framework targets functional level correctness. Although the HDL simulator is cycle-accurate, the PCIe simulation bridge and the QEMU VM are not, which causes the co-simulated time to differ from the real system run time. Table III presents a comparison between simulated time in the co-simulation framework and actual time of running the design on the physical system. Although the gap is significant and precludes performance evaluation using the co-simulation framework, the difference is acceptable for the purpose of debugging system correctness.

## V. RELATED WORK

Among prior work, the vpcie project [5] is the most similar to our co-simulation framework. Vpcie links QEMU with a VHDL simulator and is capable of full-system simulation. However, a key difference from our system is that vpcie links QEMU and HDL at a lower level. On the QEMU side, vpcie forwards low-level PCIe messages that require extra software to process, whereas our platform forwards high-level memory access and interrupt requests directly. Similarly, on the HDL side, vpcie exposes a non-standard interface with the PCIe BAR information to the FPGA platform, whereas our system uses an industry standard memory-mapped interface, reducing the complexity and improving the adaptability of our co-simulation framework.

FPGA companies and FPGA cloud vendors provide co-simulation software-HDL environments for their products and services, such as Intel OpenCL for FPGA [6], Xilinx SDAccel [7], and Amazon F1 [8]. Unlike our co-simulation framework that enables the development and debugging of the operating system and device drivers, the environments these vendors provide are limited to executing application software.

Several works use QEMU as an instruction-set simulator and connect it to virtual platforms built in SystemC for full-system co-simulation [9], [10], [11], [12]. Some of these platforms have the ability to run HDL simulations, making them similar to our work. However, these virtual platforms generally focus on ARM-based SoC ASICs using early-stage high-level hardware models rather than providing an environment to seamlessly move between production hardware and simulation. In contrast to these virtual platforms, the HDL part of our co-simulation framework is the final design that can run on the FPGA of the target system, which can be brought back into co-simulation at any time for further development and debugging.

## VI. CONCLUSIONS

In this work, we described a VM-HDL co-simulation framework for software-hardware co-design on systems with PCIe-connected FPGA boards. Our framework enables developers to have the same software, operating system, and FPGA platform running in the co-simulation environment as in the physical system, while providing much greater visibility into both the software and the hardware, and drastically reducing the debug iteration time.

## ACKNOWLEDGEMENTS

This material is based on work supported by the Semiconductor Research Corporation (SRC) and the National Science Foundation under Grant No. 1405641.